# Thermo-Poro-Mechanical Properties of Clayey Gouge and Application to Rapid Fault Shearing

J. Sulem, P. Lazar

CERMES, Institut Navier, Ecole Nationale de Ponts et Chaussées / LCPC , Marne-La-Vallée, France

I. Vardoulakis

Faculty of Applied Mathematics and Physics, National Technical University of Athens, Greece

**Abstract**

In this paper, the mechanism of fault pressurization in rapid slip events is analyzed on the basis of a complete characterization of the thermo-poro-mechanical behavior of a clayey gouge extracted at 760m depth in Aigion fault in the active seismic zone of the Gulf of Corinth, Greece. It is shown that the thermally collapsible character of this clayey gouge can be responsible for a dramatic reduction of effective stress and a full fluidization of the material. The thickness of the 'ultra localized' zone of highly strained material is a key parameter that controls the competing phenomena of pore pressure increase leading to fluidization of the fault gouge and temperature increase leading to pore fluid vaporization.

**Introduction**

The interest of the interactions between circulating fluids and fault mechanics with particular focus on the hydro-thermo-mechanical couplings has been renewed in the geophysics community during the recent years. Active fault drilling operations of the Geological Survey of Japan at Nojima Hirabayashi (Otsuki et al., 2003) and of the Corinth Rift Laboratory at Aigion (Cornet et al., 2004a) have put emphasis on the collection of core samples from the fault zones in order to study the material properties (strength, permeability, porosity) of fault gouge and the effect of pore pressure and temperature in earthquake rupture processes.

Shear heating tends to increase pore pressure and to decrease the effective compressive stress and the shearing resistance of the fault material (Lachenbruch, 1980, Mase and Smith, 1985, Wibberley and Shimamoto, 2005, Rice, 2005a,b). On the other hand dilatancy tends to decrease pore pressure. Such couplings have been studied thoroughly in a recent paper of Garagash and Rudnicki (2003). Fault zones are often characterized by large amounts of clay minerals, which form well-defined structures and orientations within the fault zone. These clay minerals inside the fault gouge are widely believed to affect significantly the mechanical behaviour of faults since as is the case for normally consolidated clays, they tend to contract when heated.

Within the frame of the 'CRL' (Corinth Rift Laboratory) project centred on the south western sector of the Gulf of Corinth (http://www.corinth-rift-lab.org), fault zone cores from the active Aigion fault have been collected continuously from depths between 708 m to 782 m. As part of the project, our work was focused on the thermo-poro-mechanical characterization of fault gouge from the Aigion well.

The Gulf of Corinth is the most seismically active zone in Europe and the fastest opening area of continental break-up, with up to 1.5 cm/year of north–south extension and more than 1 mm/year of uplift of the southern shore (Tselentis and Makropoulos, 1986). A detailed analysis of the geological structure of the studied area is presented in the recent paper of Micarelli et al. (2003). During drilling in the active Aigion fault, cuttings were collected, and geophysical logs, including borehole imaging and sonic logs, as well as vertical seismic profiles, were completed. This has provided material for a detailed lithological and structural investigation of the formations intersected by the well presented in the paper of Rettenmaier et al. (2004). The Aigion fault was intercepted at a depth of 760m, dipping at an angle of about 60°. The heart of the fault is a zone of clay-rich material derived from radiolarites about 1m long (Fig. 1). Above and below, brecciated limestones constitute the heavily damaged zone. The thickness of this zone is 3 m above the Aigion fault and 9 m below (Daniel et al., 2004).

Preliminary results from mechanical laboratory analyses on specimens taken from the Aigion fault core have been presented in previous papers (Sulem et al., 2004, 2005). In this paper, special attention is paid to the transport properties of the clayey core and temperature effects. Vardoulakis (2002a,b) has demonstrated the importance of thermally collapsible and thermally softening clay on the overall dynamic thermo-poro-mechanical behaviour of clay-rich gouges. The sensitive parameter for the description of the thermo-poro-mechanical coupling is the thermal expansion coefficient of the material. Possible collapse of the clay under thermal loading may activate fluid pressurisation inside the fault and lead to substantial reduction of the apparent friction.

The experimental data from the Aigion fault core are used in an analysis of the dynamic thermo-poro-elasto-plastic behaviour of a rapidly deforming shear band. It is shown that shear heating and fluid pressurisation is a possible mechanism leading to full fluidization of the material inside the shear band.

**Thermo-poro-mechanical characterisation of clayey gouge**

As reports from the drilling site confirm, the clayey gouge which was extracted from the Aigion fault is a remarkably soft material, (see Fig. 2). Most probably this is because this material was continuously loaded and unloaded during aseismic and seismic movements of the fault. Thus it is expected that the effect of remoulding due to sampling is of lesser importance, as long as the density and the water contents of the samples are preserved for testing. Once the Aegion fault core was taken from the borehole, it was stored for few days on site and then it was transported to the Soil Mechanics Laboratory of Ecole Nationale de Ponts et Chaussées, where it was stored in a room with controlled temperature and humidity. Obviously it was not possible to avoid partial drying of the core during extraction, storage and transportation but then it was regularly wetted as soon as it was stored in good conditions. As it was not possible to use in the experiments the same saturation fluid as in situ, we have used demineralized water in which a certain amount of particles from the sample have been placed during 24 hours. The mixture has been then filtered and the resulting fluid has been used for the saturation of the tested samples. This technique is commonly used in soil testing to avoid any chemical influence of the saturation fluid on the mechanical properties of the tested material.

The in situ density of the material is reasonably retrieved by applying the in situ stresses onto the sample. As results from direct stress measurements inside the borehole are not available at present, the in situ state of stress has to be estimated. Assuming that the density of the overburden is 2500 kg/m$^3$, we estimate a total vertical stress of about 19 MPa for a depth of 760m, which in turn corresponds to an effective vertical stress of 11.4 MPa.

**Oedometric test**

The oedometer test is commonly used in Soil Mechanics practice to measure compressibility. Here for completeness we outline the basic concepts of this test: A fluid-saturated cylindrical specimen is confined latterly by a stiff metal ring and the specimen is stressed along vertical axis. Due to the lateral rigid confinement of the specimen the strains in horizontal directions are suppressed. Porous stones are employed on the top and bottom of the specimen in order to permit the free drainage of the pore-fluid in or out of the specimen. During the test the vertical load is applied in small increments. For each load increment the vertical deformation of the specimen is monitored using displacement gauges. Assuming that the grains of the solid constituent are incompressible all volume changes are attributed to changes in the voids ratio of the specimen and are directly linked with the extrusion of the pore-fluid from the pore-space. Thus we relate the change in the voids ratio of the specimen $\Delta e$ with the volumetric strain $\varepsilon_v$:

$$\varepsilon_v = \frac{\Delta e}{1 + e_0} \qquad (1)$$

where $e_0$ is the initial voids ratio.

The commonly observed non-linear stress-strain response of a soft geomaterial in oedometric compression makes it more convenient to plot the void ratio versus the effective vertical stress $\sigma'_v$ in a logarithmic scale because. As was first observed by Terzaghi (1943), at large stresses the corresponding curve becomes a straight line, so that a unique parameter, the so-called compression index

$$C_c = -\frac{\Delta e}{\Delta(\log \sigma'_v)} \tag{2}$$

is sufficient to characterize the compressibility of the material over a wide range of stresses. Similarly one is defining upon unloading the so-called swelling index $C_s$.

In Fig. 2a we see the results from oedometric compression tests performed on a saturated sample using a high-pressure oedometer The measured compression index $C_c=0.11$ is typical of a sandy clay. The oedometric test allows the determination of the isothermal compressibility coefficient.

$$c = \frac{1}{1+e} \frac{\overline{C}_c}{|\sigma'|}, \quad \overline{C}_c = \frac{C_c}{\ln(10)} \tag{3}$$

Thus for loading at 11.4 MPa, the isothermal, compressibility coefficient is estimated as $c \approx 0.0039 \text{ MPa}^{-1}$.

By applying the principles of the consolidation theory commonly used in geotechnical engineering (Terzaghi, 1943), it is possible at each stage of loading to evaluate the one-dimensional consolidation coefficient $c_v$ using Taylor or Casagrande methods. The hydraulic conductivity of the gouge is evaluated as a function of the applied oedometric pressure (Fig. 2b)

$$K_w = c_v \frac{\rho_w g}{1+e} \frac{\Delta e}{\Delta \sigma'_v} \qquad (4)$$

where e is the actual voids ratio and $\Delta e$ its increment.

Accordingly the consolidation coefficient and the hydraulic conductivity at 11.4 MPa effective normal stress are $c_v \approx 2.8 \times 10^{-8} \, m^2/s$ and $K_w \approx 10^{-12} \, m/s$. The corresponding physical permeability is $k_w \approx 10^{-19} \, m^2$. In direct permeability tests performed by pumping degassed-deionised water on a sample of the red clay material Song et al. (2004) have obtained ranges of 0.9 to $2 \times 10^{-18}$ $m^2$. This result is consistent with those published in the literature. For example for the Nojima fault, Lockner et al. (2000) obtain a permeability of about $10^{-19}$ $m^2$ in the center of the shear zone; the same order of magnitude is obtained by Wibberley and Shimamoto (2003) for the Median Tectonic Line Fault, Japan.

**Strength parameters from drained triaxial compression tests**

Drained triaxial compression tests have been performed at room temperature (22°C) and at 70°C, and at 8, 16 and 18 MPa confining pressure, on samples of the clayey gouge. The corresponding stress-strain curves are shown on Fig. 3a,b.

At room temperature, for the test at 8 MPa confining pressure, the loading phase was interrupted at an axial strain of 7.2 % corresponding to a deviatoric stress of 9.9 MPa. This deviatoric stress is actually less than the stress at which failure occurs. Thus the same sample could be used for another experiment if loaded at a different higher pressure. For the tests at 16 MPa and 18 MPa, the material was compacting and reached the critical state (i.e. for zero dilatancy) for a maximum friction angle of $\phi=27.9°$. In triaxial compression the maximum friction angle is computed from the peak stress of the loading curve as (Vardoulakis & Sulem 1995)

$$\sin\phi = \frac{3\tan\psi}{2\sqrt{3}+\tan\psi} \quad \text{with} \quad \tan\psi = \frac{|\sigma_r - \sigma_z|/\sqrt{3}}{(2\sigma_r + \sigma_z)/3} \qquad (5)$$

where $\sigma_r$ is the radial stress (confining pressure) and $\sigma_z$ is the maximum axial stress. The dilatant behaviour which occurs at the end of the test at 18 MPa of confinement is attributed to the occurrence of shear-banding in the sample.

The frictional resistance at critical state of the material increases slightly with temperature and strain rate. The triaxial experiment run at 70°C at the same strain rate as the ones at room temperature ($10^{-6}\,\text{s}^{-1}$) yielded a maximum friction angle of 29°.

As shown in Fig. 4b changes in the strain rate during the test show also some rate sensitivity for the frictional resistance, with deviatoric stress increasing with an increased strain rate and decreasing with a decreased strain rate.

**Drained thermal loading of the clayey gouge**

In order to study the effect of temperature on the volumetric behaviour of the material, two heating tests of drained specimens have been performed under isotropic stress conditions. For each test, the sample is first loaded isotropically to its final stress state in drained conditions and at constant room temperature. Then, keeping the isotropic stress constant, the sample was heated in drained conditions at a rate of 0.02°C/minute up to a maximum of temperature change of 32°C. The experimental results are shown in Fig. 4.

The two tests show a good reproducibility and show also that the final state of isotropic stress is not influencing the response.

The important result from these experiments is that the clay-rich material is contracting when heated. This phenomenon represents a thermo-plastic or structural collapse. The corresponding elasto-plastic contraction coefficient is negative $\alpha = -2.4 \times 10^{-4} / °C$. During the cooling phase thermo-elastic contraction is occurring at a rate of $\alpha^e = 1.4 \times 10^{-4} / °C$. The resulting isobaric thermo-plastic contraction coefficient is thus $\alpha_c^p = \alpha - \alpha^e = -3.8 \times 10^{-4} / °C$.

The thermal pressurisation $\Lambda$ in undrained conditions is evaluated as (see appendix A)

$$\Lambda = -\frac{(\alpha - \alpha^e)}{c} \qquad (6)$$

The numerical estimation of $\Lambda$ is 0.1MPa/°C.

**Application to shear heating and fluid pressurization during seismic slip**

We consider a rapidly deforming and long shear band of thickness $d_B$ consisting of water-saturated material soil. Inside such a shear-band the pore-pressure $p(t,z)$, the temperature $T(t,z)$ and the velocity $v(t,z)$ are assumed to be functions only of time t and of position z in the direction normal to the band (Figure 5).

**Governing equations**

The equations that govern the problem of dynamic, thermo-poro-plastic shearing can be derived from the corresponding balance laws for mass, momentum and energy for a two-phase saturated porous medium.

*Mass balance*

Conservation of fluid mass is expressed by

$$\frac{\partial m}{\partial t} + \frac{\partial q_f}{\partial z} = 0 \tag{7}$$

where m is the fluid mass per unit volume of porous medium (in the reference state) ($m = \rho_w n$) and $q_f$ is the flux of fluid

The first term of equation (7) is evaluated as (see Appendix 1)

$$\frac{\partial m}{\partial t} = n\rho_w (\beta_n + \beta_w)\frac{\partial p}{\partial t} + \rho_w (\alpha - (1-n)\alpha_s - n\lambda_w)\frac{\partial T}{\partial t} \tag{8}$$

where

- $\rho_w$ is the is the density of water,
- n is the volume of free water per unit volume of porous medium (in the reference state) (pore fluid volume fraction),
- $\beta_w$ and $\lambda_w$ are respectively the compressibility and the thermal expansion coefficient of the pore water (free water)
- $\beta_n$ and $\lambda_n$ are respectively the pore volume compressibility and thermal expansion coefficient of the pore volume.

- $\alpha$ is the drained thermal expansion coefficient of the saturated soil and
- $\alpha_\sigma$ is the thermal expansion coefficient of the solid fraction.

The second term of equation (7) is evaluated assuming Darcy's law

$$q_f = -\frac{\rho_w}{\eta_w} k_w \frac{\partial p}{\partial z} \qquad (9)$$

In equation (9) $k_w$ is the permeability and $\eta_w$ is the fluid viscosity.

Substituting (8) and (9) into (7) gives the fluid mass conservation equation

$$\frac{\partial p}{\partial t} = \Lambda \frac{\partial T}{\partial t} + \frac{1}{n\rho_w(\beta_n + \beta_w)} \frac{\partial}{\partial z}\left(\rho_w \frac{k_w}{\eta_w} \frac{\partial p}{\partial z}\right) \qquad (10)$$

where

$$\Lambda = \frac{n\lambda_w - (\alpha - (1-n)\alpha_s)}{n(\beta_w + \beta_n)} \qquad (11)$$

*Energy balance equation*

In a similar manner conservation of energy is expressed as

$$\rho C \frac{\partial T}{\partial t} + \frac{\partial q_h}{\partial z} - \Psi_p = 0 \qquad (12)$$

where ρC is the specific heat per unit volume of the fault gouge in its reference state, $q_h$ is the heat flux and $\Psi_p$ is the rate of mechanical energy dissipation due to inelastic deformation.

The heat flux is related to the temperature gradient by Fourier's law

$$q_h = -k_T \frac{\partial T}{\partial z} \qquad (13)$$

where $k_T$ is the thermal conductivity of the saturated material. In equation (12) it is assumed that all heat flux is due to heat conduction neglecting heat convection by the moving hot fluid. This assumption is justified by the low pore volume fraction and the low permeability of fault gouge.

If we neglect all dissipation in the fluid the rate of mechanical energy dissipation is written as

$$\Psi_p = \tau \frac{\partial v}{\partial z} \qquad (14)$$

where v is the local fault parallel velocity and τ is the shear stress. In equation (8) the work done by the normal stress $\sigma_n$ is considered as negligible as compared to the one done by τ at the large shear considered.

Substituting (13) and (14) into (12) gives the energy conservation equation

$$\frac{\partial T}{\partial t} = \frac{1}{\rho C} \frac{\partial}{\partial z}\left(k_T \frac{\partial T}{\partial z}\right) + \frac{1}{\rho C} \tau \frac{\partial v}{\partial z} \qquad (15)$$

*Momentum balance equation*

The momentum balance equation is restricted in this study to the one-dimensional formulation in the z-direction as the length scales in the direction parallel to the fault over which the mechanical fields vary are much larger than in the direction normal to it. The 1D-momentum balance equation reads as

$$\frac{\partial \tau}{\partial z} = \rho \frac{\partial v}{\partial t} \qquad (16)$$

As discussed by Rice (2006), the effect of even large accelerations like several times the acceleration of the gravity *g* is insignificant over the small length scales in the z-direction normal to the fault where the heat and fluid diffusion process are taking place during rapid slip and very high values of pressure and temperature gradients. For example, assuming an acceleration of 10*g* and a specific mass of the material of 2500 kg/m$^3$ would result in a change of $2.5 \times 10^{-2}$ MPa/m for $\tau$. The relevant length scale in z-direction is only few mm and thus the variation for $\tau$ can be neglected and mechanical equilibrium can be assumed.

$$\frac{\partial \tau}{\partial z} = 0 \qquad (17)$$

Consequently, as the shear stress is constant in space, the Coulomb friction law cannot be assumed to be met in all deforming regions unless the pore-pressure is also constant in space as it is the case in the undrained adiabatic limit.

It is thus assumed that the frictional resistance is proportional to the mean effective stress inside the band

$$\tau(t) = f\left(\sigma_n - \frac{1}{h}\int_{-h/2}^{h/2} p(\xi, t)d\xi\right) \qquad (18)$$

where f is the friction coefficient of the gouge and h is the width of the shear band.

**The undrained adiabatic limit**

A significant simplification of the mathematical model arises if one neglects the diffusion of heat in the surroundings of the rapidly deforming shear-band. This case corresponds to the so-called adiabatic limit. The model may be simplified further if one neglects pore-pressure diffusion as well, which corresponds to the 'undrained' limit. The undrained adiabatic limit is applicable as soon as the slip event is sufficiently rapid and the shear zone broad enough to effectively preclude heat or fluid transfer. Another idealized situation is met if one assumes that the shear zone is infinitely small. This situation corresponds to the model of slip on a plane for which an analytical solution has been proposed by Rice (2005, 2006). These two extreme cases provide good first approximations of the phenomenon. However, as shown by Rice (2006), the solution of slip on a plane overestimates the temperature field, and on the other hand, the adiabatic and undrained solution appears to be unstable for localization.

The set of coupled governing equations for adiabatic, undrained shear for $0 \leq z \leq h/2$ and $t > 0$ reads as follows

$$\text{mass balance}: \frac{\partial p}{\partial t} = \Lambda \frac{\partial T}{\partial t} \tag{19}$$

$$\text{energy balance}: \frac{\partial T}{\partial t} = \frac{\tau}{\rho C} \frac{\partial v}{\partial z} = \frac{(\sigma_v - p)f}{\rho C} \frac{\partial v}{\partial z}$$

where $\sigma_v$ is the geostatic pressure

In the undrained adiabatic limit, p is independent of z (i.e. $\frac{\partial p}{\partial z} = 0$). From the equation of momentum balance we obtain a linear distribution for the velocity field corresponding to homogeneous straining:

$$v = v_{slip} \frac{z}{h/2} \tag{20}$$

The solution is designed for constant velocity. An analytic solution can then be obtained under the following form:

$$p = p_0 + (\sigma_v - p_0)\left(1 - \exp(-\beta \frac{u}{h/2})\right) \tag{21}$$

$$T = T_0 + \frac{(\sigma_v - p_0)}{\Lambda}\left(1 - \exp(-\beta \frac{u}{h/2})\right)$$

where $p_0$ is the initial hydrostatic pore pressure, u is the accumulated slip displacement ($u = \int v_{slip} dt$) and $\beta = \frac{\Lambda}{\rho C} f$.

This means that the pore-pressure increases towards its geostatic limit $\sigma_v$ which corresponds to full fluidization exponentially with the slip displacement. In due course of the shear heating and fluid pressurization process, the shear strength $\tau$ is thus reduced towards zero. The governing physical factor is the coefficient $\beta$, which combines the effects of friction and temperature-pressure coupling and depends also on the specific heat of the mixture. For the considered gouge $\rho C \approx 2.7$ MPa/°C, f=0.5 and $\Lambda$=0.1 MPa/°C which gives $\beta = 0.0185$. The temperature rise and the pore pressure rise are simply linked by the thermal pressurization coefficient $\Lambda$

$$T - T_0 = \frac{1}{\Lambda}(p - p_0) \qquad (22)$$

The results of the evolution of the temperature and of the effective stress are shown on Fig. 6 as a function of the normalized displacement $u/(h/2)$ for a slip velocity of 1m/s. Assuming that the density of the overburden is 2500 kg/m$^3$, we estimate a total vertical stress of about 19 MPa for a depth of 760m, which in turn corresponds to an effective vertical stress of 11.4 MPa.

A key parameter is the actual shear-band thickness. It appears that in the clayey core of the fault zone, an 'ultra-localized' zone of highly strained material is formed. The thickness of this zone is of the order of tenths of a millimetre despite a much wider fault zone (Cocco and Rice, 2002). Thus slip at a centimetre scale can actually induce fluidization of the material inside the shear-band.

**Effect of fluid and heat diffusion**

Rice (2005) has shown that the above homogeneous undrained adiabatic solution is actually unstable in the sense that small spatial non-uniformities with wave lengths greater than a critical value grow exponentially with time. For the typical values of the thermo-poro-mechanical parameters of the considered gouge, it can be shown that this critical wave length is of the order of few tens of mm which is comparable to the observed thickness of shear zones where individual slip events actually localize. After localization, heat and fluid transfer cannot be neglected.

The coupled diffusion equations which can be summarized as:

$$\frac{\partial p}{\partial t} = \Lambda \frac{\partial T}{\partial t} + \frac{1}{n(\beta_n + \beta_w)} \frac{\partial}{\partial z}\left(\frac{k_w}{\eta_w}\frac{\partial p}{\partial z}\right) \tag{23}$$

$$\frac{\partial T}{\partial t} = \frac{1}{\rho C}\frac{\partial}{\partial z}\left(k_T \frac{\partial T}{\partial z}\right) + \frac{1}{\rho C} f\left(\sigma_n - \frac{1}{h}\int_0^h p(\xi,t)d\xi\right)\frac{\partial v}{\partial z}$$

The source term for the temperature field decreases with increasing pore pressure. This system shows the competing effect of pore fluid pressurization and temperature increase. In the following we will examine some numerical examples for which fluidization the clayey gouge or vaporization of the pore fluid may occur.

This system is solved numerically using an explicit 2$^{nd}$ order Runge-Kutta finite difference scheme.

$c_{th} = \frac{k_T}{\rho C}$ is the thermal diffusivity of the saturated soil which value is 0.2mm$^2$/s for the considered material and $c_{hy} = \frac{k_w}{n(\beta_n + \beta_w)\eta_w} = \frac{k_w}{(c + n\beta_w)\eta_w} \approx \frac{k_w}{c\eta_w}$ is the hydraulic diffusivity which value is $c_{hy}$=0.026mm$^2$/s at 20°C and at atmospheric pressure. However, we have to account for the temperature dependency of the viscosity of water and the hydraulic diffusivity is approximated with the following function:

$$c_{hy}\ (\text{in mm}^2/s) = 9.4661\times10^{-4}\,T(\text{in °C}) + 1.2589\times10^{-3} \tag{24}$$

The thermal collapse of the clayey gouge occurs only inside the shear band. Outside the band the material is unloaded elastically:

$$\text{for } |z| \leq h/2, \Lambda \approx \frac{n(\lambda_w - \alpha_s) - (\alpha - \alpha_s)}{c} \approx \frac{-(\alpha - \alpha_s)}{c} \approx 0.1 \text{MPa} \quad (25)$$
$$\text{for } |z| > h/2, \Lambda \approx 0$$

Note that for a rock in the elastic range, the pressurization coefficient is given by equation (A16) in Appendix 1. With typical values of compressibility and thermal expansion coefficient of rock and water at great depth, it is obtained that the 'elastic' pressurization coefficient is not negligible and has values between 0.3 and 1 (Rice 2006). For a clayey gouge at relatively low depth (760m), thermal collapse is the only mechanism of fluid pressurization.

Far field boundary conditions correspond to the initial temperature of 28°C and total stress of 19 MPa.

The numerical results are shown on Fig. 7 assuming a shear band thickness 2h=1mm. The liquefaction limit ($p(0,t) = \sigma_v$, i.e. full fluidization of the gouge), is reached after 0.13s of shearing. We observe that, as expected, the increase of temperature is more important than in case of undrained adiabatic shearing.

If we assume a shear band thickness which is twice smaller (h=0.5mm), then computed results show that vaporization of the pore fluid occurs (Fig. 8). This state is reached at t=1.4s for which p=17MPa and T=356°C. Note that fluid vaporization is possible only if it is assumed that the soil is not fully saturated.

We would like to emphasize here that in these computations the coefficient of pressurization $\Lambda$ was assumed to be constant. We have no experimental data about the thermal collapse of clayey soils at high temperatures. We might expect that the pressurization coefficient will decrease with increasing temperature due to the limited amount of structural water that can be released when heated.

**Conclusion**

In rapid fault shearing of clay-rich faults the effects of pore-water heating and consequent pore-pressure rise cannot be disregarded. The Aigion fault gouge, where extracted from the Aigion fault at a depth of 760 m, is a clay-rock mixture. The experimental characterization program performed on cores of this material demonstrates that although the clay fraction is relatively small, it has a significant influence on the thermo-mechanical properties of the material. The clayey core of the fault has a very low fluid permeability and exhibits <u>contractant</u> volumetric behaviour when heated. The sensitive parameter for the description of the thermo-poro-mechanical coupling is the thermal pressurisation coefficient of the material. It was found that at a nominal effective stress corresponding to 760 m deep conditions, this coefficient is about 0.1 MPa/°C. An analysis of undrained and adiabatic shearing of a thin layer of this clayey gouge has shown that shear heating and fluid pressurization is a possible mechanism leading to full fluidization of the material inside the shear band. A numerical solution including heat and fluid diffusion has also been proposed. It has been shown that the thickness of the shear band is a key parameter for the competing effects of fluidization of the gouge or vaporization of the pore fluid.

We emphasize the fact that the parameters used in the analysis are the ones of a fault gouge at 760 m depth. The actual nature and rheology of materials at seismogenic depth is a challenging question that can only be answered in the future by ultra-deep drilling.

**Acknowledgements**

The authors wish to acknowledge the EU projects *DG-Lab Corinth* (EVR1-CT-2000-40005) and *Fault, Fractures and Fluids: Golf of Corinth*, in the framework of program Energy (ENK6-2000-0056) and the 'Groupement de Recherche Corinthe' (GDR 2343), INSU-CNRS for supporting this research. The authors wish to thank James Rice for the discussions held during the trimester on granular materials at Institut Poincaré in Paris, 2005. They wish also to thank François Martineau (LCPC) for his assistance in the experimental work.

**Appendix 1**

**Thermal volume changes of saturated clay soils**

When temperature is increased in drained conditions, a normally consolidated clayey soil exhibits an irreversible contraction. This thermal contraction is related to physicochemical clay-water interactions based on the change in thickness of the double layer with an increase in temperature promoting eventually a breakdown of the adsorbed water. In undrained conditions the pore pressure may develop as the result of heating. Experimental results reported by Hueckel and Pellegrini, 1991, have shown the failure of samples heated from 70 to 90°C under undrained conditions with a constant stress deviator.

On the phenomenological level the thermal strain of normally consolidated clays can be considered with the frame of thermoplastic hardening theory (e.g. Baldi et al. 1988, Hueckel and Baldi, 1990, Sultan et al. 2002.)

The deformation of the clay skeleton can be described using a thermoelastic law. The collapse of adsorbed water results in the rearrangement of the internal structure of the soil skeleton which causes an irreversible volume reduction which can describe using thermoplasticity.

The porous space of a clay can be seen as a double porosity network. The fluid mass m per unit volume of porous medium (in the reference state) can be written as the sum of the mass of the free water $m_f$ and the mass of the 'structural' water $m_a$. The structural water is bonded to the solid mineral surfaces by physico-chemical forces. The density and the thermal expansion of the structural water are different from that of bulk water.

$$m = m_f + m_a \tag{A1}$$

$$m_f = \rho_w n \tag{A2}$$

where $\rho_w$ is the is the density of water and n is the volume of free water per unit volume of porous medium (in the reference state) (pore fluid volume fraction).

$$n = \frac{V_{pore}}{V_{tot,0}} \tag{A3}$$

n is essentially the porosity $\phi = V_{pore}/V_{tot}$ if $V_{tot}$ is nearly equal to its reference-state value $V_{tot,0}$. n is also called the Lagrangian porosity as opposed to the Eulerian porosity $\phi$ (Coussy, 2004).

The structural water occupies a volume $V_a$ and the structural water volume fraction $n_a$ is

$$n_a = \frac{V_a}{V_{tot,0}} \tag{A4}$$

The increment of pore fluid volume fraction dn is decomposed into a reversible part $dn^e$ which is expressed in terms of increments dp of pore pressure and dT of temperature and an irreversible part $dn^p$ describing the thermal collapse of clay

$$\begin{aligned} dm_f &= n d\rho_w + \rho_w dn \\ d\rho_w &= \rho_w \beta_w dp - \rho_w \lambda_w dT \\ dn &= n\beta_n dp + n\lambda_n dT + dn^p \end{aligned} \tag{A5}$$

where $\beta_w$ and $\lambda_w$ are respectively the compressibility and the thermal expansion coefficient of the pore water (free water) and $\beta_n$ and $\lambda_n$ are respectively the (elastic) pore volume compressibility and thermal expansion coefficient of the pore volume.

$$dm = n\rho_w\left((\beta_n + \beta_w)dp + (\lambda_n - \lambda_w)dT\right) + \rho_w dn^p + dm_a \qquad (A6)$$

The variation of the mass of the structural water $dm_a$ corresponds to the quantity of structural water transformed into free water during the heating process ($dm_a \leq 0$).

In a drained heating test of a saturated sample under constant isotropic total stress, one can measure independently the incremental volumetric strain of the sample $d\varepsilon_v$ and the incremental volume of water expelled from the saturated sample during drained heating $dV_{dr}$.

The change of the mass of fluid per unit volume (in the reference state) in a drained test is obtained from equation (A5) with $dp = 0$

$$dm = n\rho_w(\lambda_n - \lambda_w)dT + \rho_w dn^p + dm_a \qquad (A7)$$

This incremental mass is equal (with opposite sign) to the incremental mass of the drained fluid $dm_{dr}$

$$dm = -dm_{dr} = -\rho_w \frac{dV_{dr}}{V_{tot,0}} \qquad (A8)$$

The incremental volume of drained water is the difference between the water volume change and the volume change of the total porous space.

$$dV_{dr} = dV_w - dV_{pore} - dV_a \tag{A9}$$

The water volume change corresponds to the thermal expansion of the free water.

$$dV_w = V_w \lambda_w dT \tag{A10}$$

The volume change of the total porous space

$$\frac{dV_{pore} + dV_a}{V_{tot,0}} = \frac{dV_{tot}}{V_{tot,0}} - \frac{dV_s}{V_{tot,0}} = d\varepsilon_v - (1-n)\alpha_s dT \tag{A11}$$
$$d\varepsilon_v = \alpha dT$$

In equation (A11) $\alpha$ is the drained thermal expansion coefficient of the saturated soil and $\alpha_s$ is the thermal expansion coefficient of the solid fraction. This coefficient can be positive or negative depending on the over-consolidation ratio of the clayey soil.

From equations (A8-A10) we deduce

$$\frac{dV_{dr}}{V_{tot,0}} = n\lambda_w dT - \left(\alpha - (1-n)\alpha_s\right)dT \tag{A12}$$

Consequently using equations (A7) and (A8) we obtain the following expression of the incremental change of mass of the structural water

$$dm_a = \rho_w\left(\alpha - (1-n)\alpha_s - n\lambda_n\right)dT - \rho_w dn^p \qquad (A13)$$

In an undrained heating test the pore fluid pressurization can be deduced from equations (A7) and (A13) (with dm = 0) assuming that the mass of structural water released in the heating process is not affected by the drainage conditions. This assumption is consistent with the experimental observation that the volumetric strain during heating of normally consolidated clay is not affected by the stress level.

$$dp = \Lambda dT \qquad (A14)$$
$$\text{with } \Lambda = \frac{n\lambda_w - \left(\alpha - (1-n)\alpha_s\right)}{n(\beta_w + \beta_n)}$$

In case of purely thermo-poro-<u>elastic</u> material

$$\alpha = \alpha^e = \alpha_s \qquad (A15)$$

The expression of the elastic pressurization coefficient given by Rice (2006) is thus retrieved

$$\Lambda = \frac{\lambda_w - \lambda_n}{\beta_w + \beta_n} \qquad (A16)$$

The elastic pore volume compressibility $\beta_n$ can be expressed in terms of the compressibility of the material c and of the solid grains $c_s$ using the equations of linear thermo-poro-elasticity (Mc Tigue, 1986, Rice, 2006)

$$dn^e = (c - c_s)\left(\frac{d\sigma_{kk}}{3} + dp\right) - nc_s dp + n\alpha_s dT \qquad (A17)$$

Under constant total stress ($d\sigma_{kk} = 0$) comparing equations (17) and (5) identifies

$$\beta_n = (c - (1+n)c_s)/n \qquad (A18)$$
$$\lambda_n = \alpha_s$$

For a clayey gouge at relatively low depth $c_s \ll c$ and consequently $\beta_n \approx c/n \gg \beta_w$. Consequently the pressurization coefficient is expressed as

$$\Lambda \approx \frac{n(\lambda_w - \alpha_s) - (\alpha - \alpha_s)}{c} \approx \frac{-(\alpha - \alpha_s)}{c} \qquad (A19)$$

**FIGURE CAPTIONS**

**Fig. 1.** : Box 49 containing the core at depth 759.70 m, characterized as "Aigion fault" core

**Fig. 2.** : Oedometric compression tests: (a) Oedometric curve; (b) Porosity and permeability evolution

**Fig. 3**. Drained triaxial tests : (a) at room temperature (22°C), (b) at 70°C

**Fig. 4.** Drained heating tests under isotropic state of stress

**Fig. 5.** The deforming shear-band with heat and fluid fluxes

**Fig. 6.** Undrained adiabatic shear-heating and fluid pressurization

**Fig. 7** (a) Temperature and pore pressure vs time at the center of the band; (b) Temperature and pore pressure field at t=0.13 (liquefaction)  (Shear band thickness=1mm – slip velocity=1m/s)

**Fig. 8.** Temperature and pore pressure vs time at the center of the band. Vaporization of the pore fluid is reached at 1.4s (Shear band thickness=0.5mm – slip velocity=1m/s)

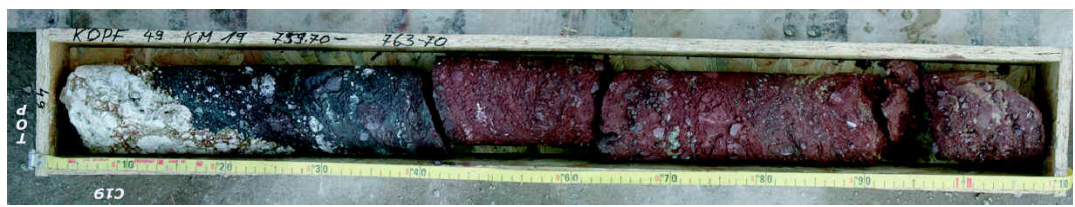

**Fig. 1.** : Box 49 containing the core at depth 759.70 m, characterized as "Aigion fault" core

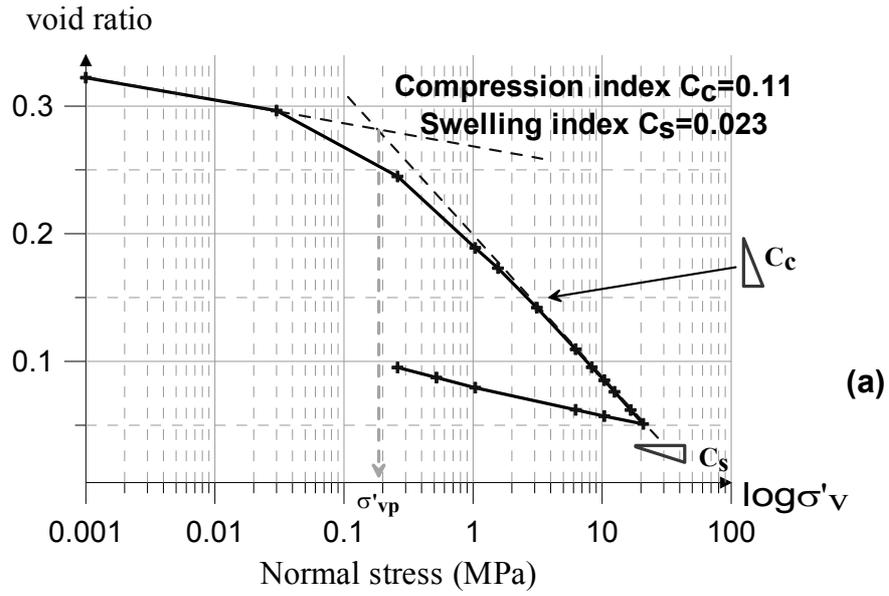

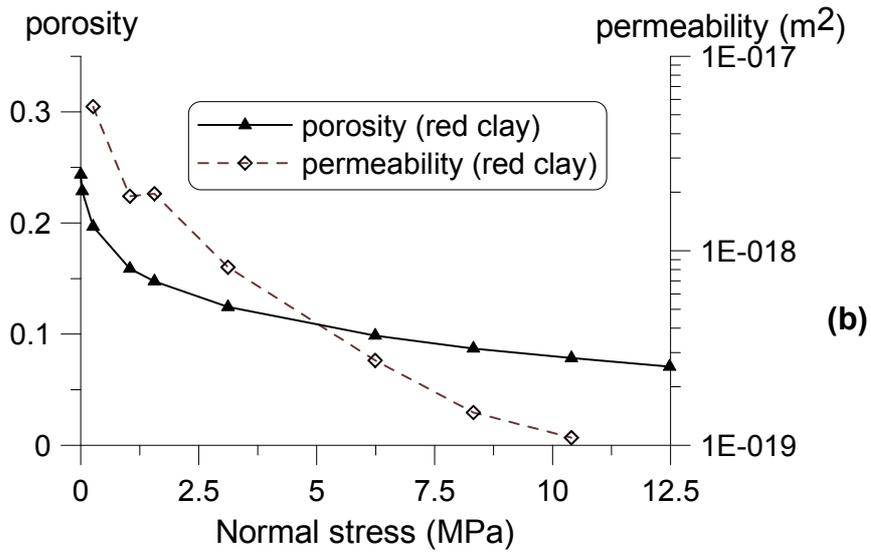

**Fig. 2.** : Oedometric compression tests: (a) Oedometric curve; (b) Porosity and permeability evolution

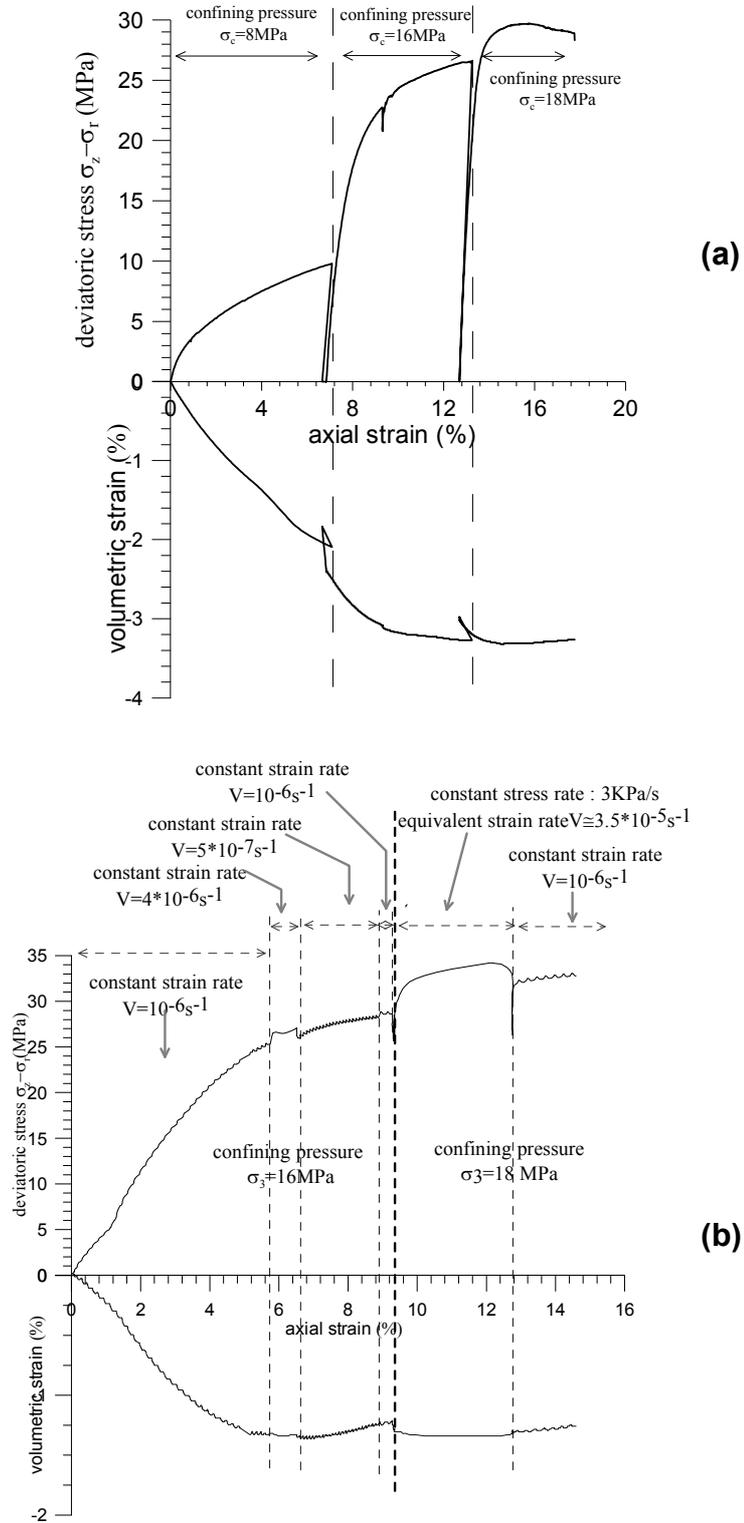

**Fig. 3**. Drained triaxial tests : (a) at room temperature (22°C), (b) at 70°C

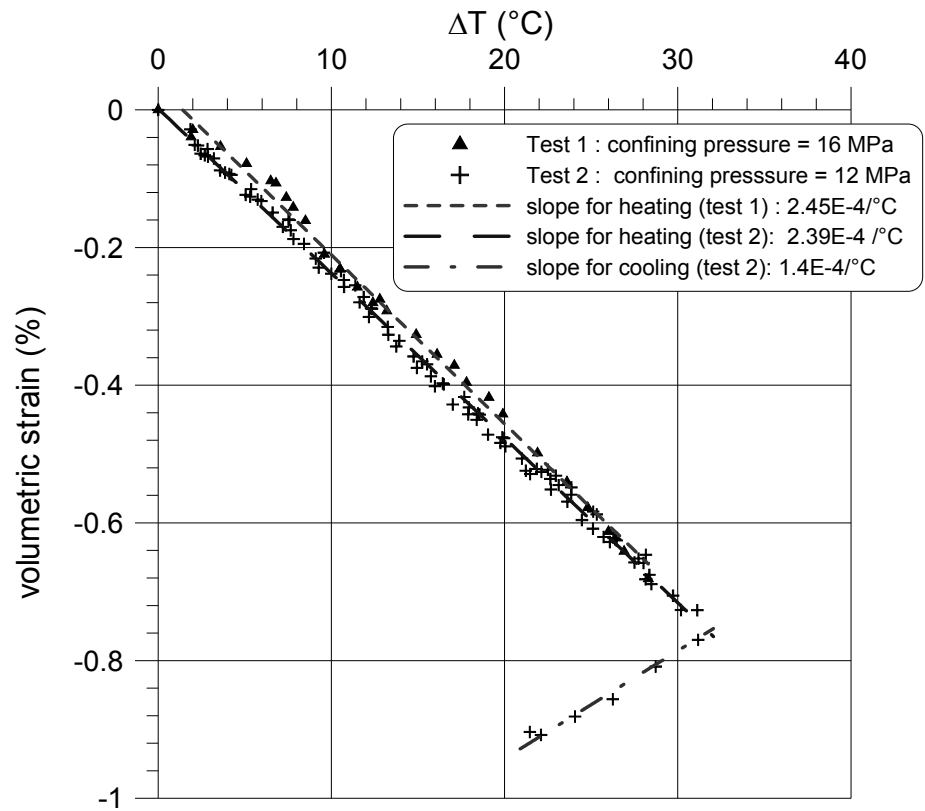

**Fig. 4.** Drained heating tests under isotropic state of stress

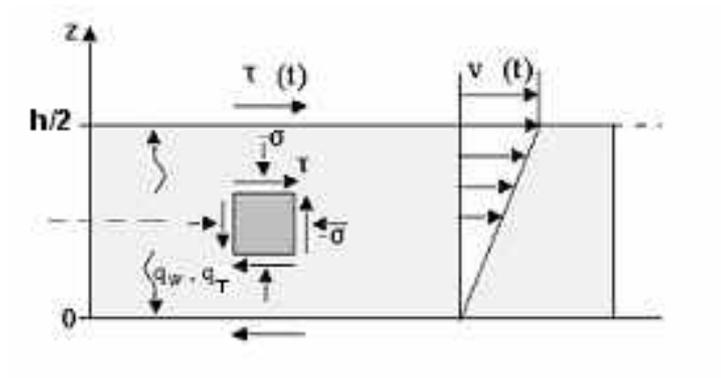

**Fig. 5.** The deforming shear-band with heat and fluid fluxes

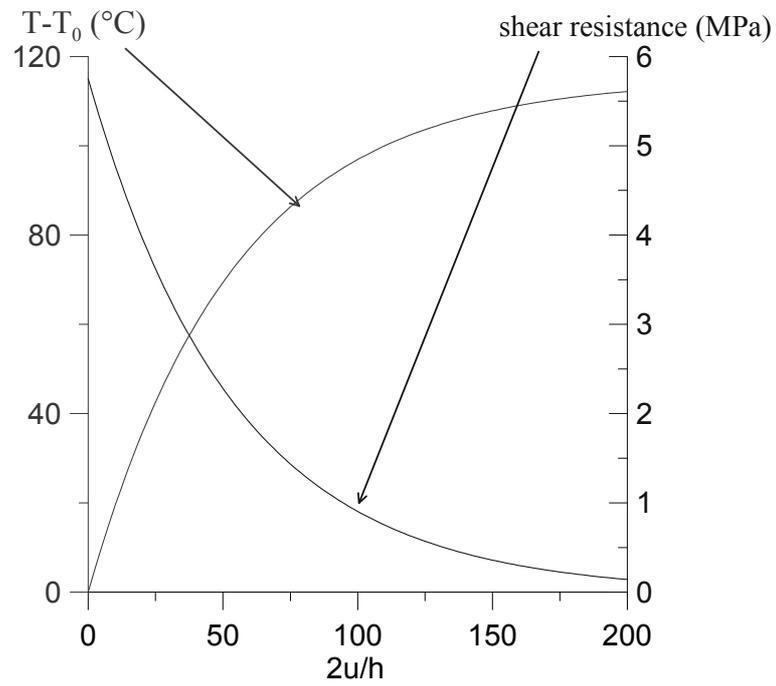

**Fig. 6.** Undrained adiabatic shear-heating and fluid pressurization

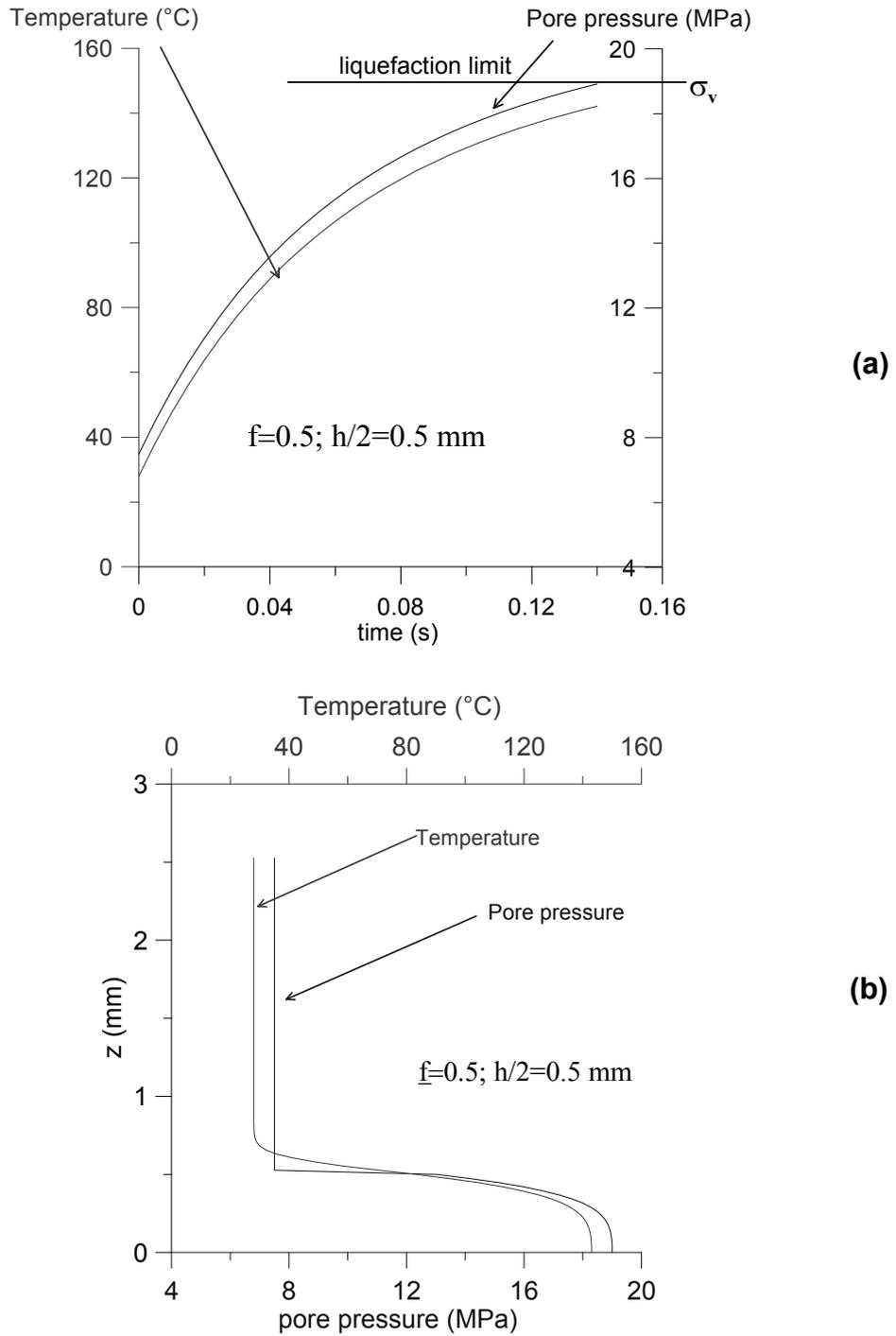

**Fig. 7** (a) Temperature and pore pressure vs time at the center of the band; (b) Temperature and pore pressure field at t=0.13 (liquefaction) (Shear band thickness=1mm – slip velocity=1m/s)

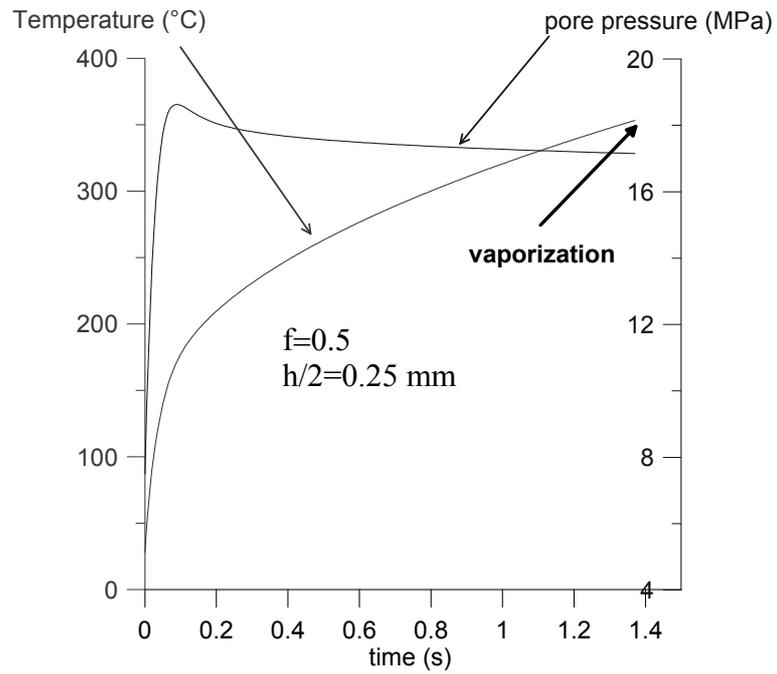

**Fig. 8.** Temperature and pore pressure vs time at the center of the band. Vaporization of the pore fluid is reached at 1.4s (Shear band thickness=0.5mm – slip velocity=1m/s)